\documentclass[twoside]{article}
\usepackage{Proc_NTSE}
\usepackage{amsmath}

\graphicspath{{figures/}}

\newcommand{\beq}[0]{\begin{equation}}
\newcommand{\eeq}[0]{\end{equation}}
\newcommand{\beqfn}[0]{\begin{footnotesize}\[}
\newcommand{\eeqfn}[0]{\]\end{footnotesize}}

\newcommand{\fmi}{\mbox{\,fm}^{-1}}

\newcommand{\Uop}{\wh U}
\newcommand{\Uopdag}{\wh U^{\dagger}}
\newcommand{\wh}[0]{\widehat}

\newcommand{\wt}{\widetilde}
\newcommand{\Uwh}{\widehat U}

\newcommand{\qvec}{{\bf q}}
\newcommand{\kvec}{{\bf k}}

\newcommand{\adag}{a^\dagger}

\newcommand{\ap}{a^{\phantom{\dagger}}}

\newcommand{\flow}{\lambda}
\newcommand{\la}{\langle}
\newcommand{\ra}{\rangle}

\newcommand{\I}{\item}

\newcommand{\kl}{{k}}
\newcommand{\kpl}{{k'}}
\newcommand{\qh}{{q}}
\newcommand{\qph}{{q'}}

\newcommand{\vlowk}{V_{{\rm low\,}k}}
\newcommand{\Ohat}{{\widehat O}}

\begin{document}
\thispagestyle{plain}

\begin{center}
{\Large \bf \strut
 High-resolution probes of low-resolution nuclei
\strut}\\
\vspace{10mm}
{\large \bf 
R.J.\ Furnstahl$^{a}$}
\end{center}

\noindent{
\small $^a$\it Department of Physics, Ohio State University, Columbus, OH\ \ 43085} 

\markboth{
R.J.\ Furnstahl}
{
High-resolution probes of low-resolution nuclei} 

\begin{abstract}
Renormalization group (RG) methods used to soften Hamiltonians allow large-scale computational resources to be used to greater advantage in calculations of nuclear structure and reactions. 
These RG transformations lower the effective resolution of the nuclei, which raises
questions about how to calculate and interpret high-momentum transfer probes of nuclear structure. 
Such experiments are conventionally explained in terms of short-range correlations, but 
these disappear with the evolution to low-momentum scales.
We highlight the important issues and prospects in the context of 
recent developments in RG technology, with guidance from  
the analogous extraction of parton distributions.
\\[\baselineskip] 
\end{abstract}

\section{Introduction}  \label{sec:intro}

\begin{figure}[hb!]
  \centerline{\includegraphics[width=0.5\textwidth]{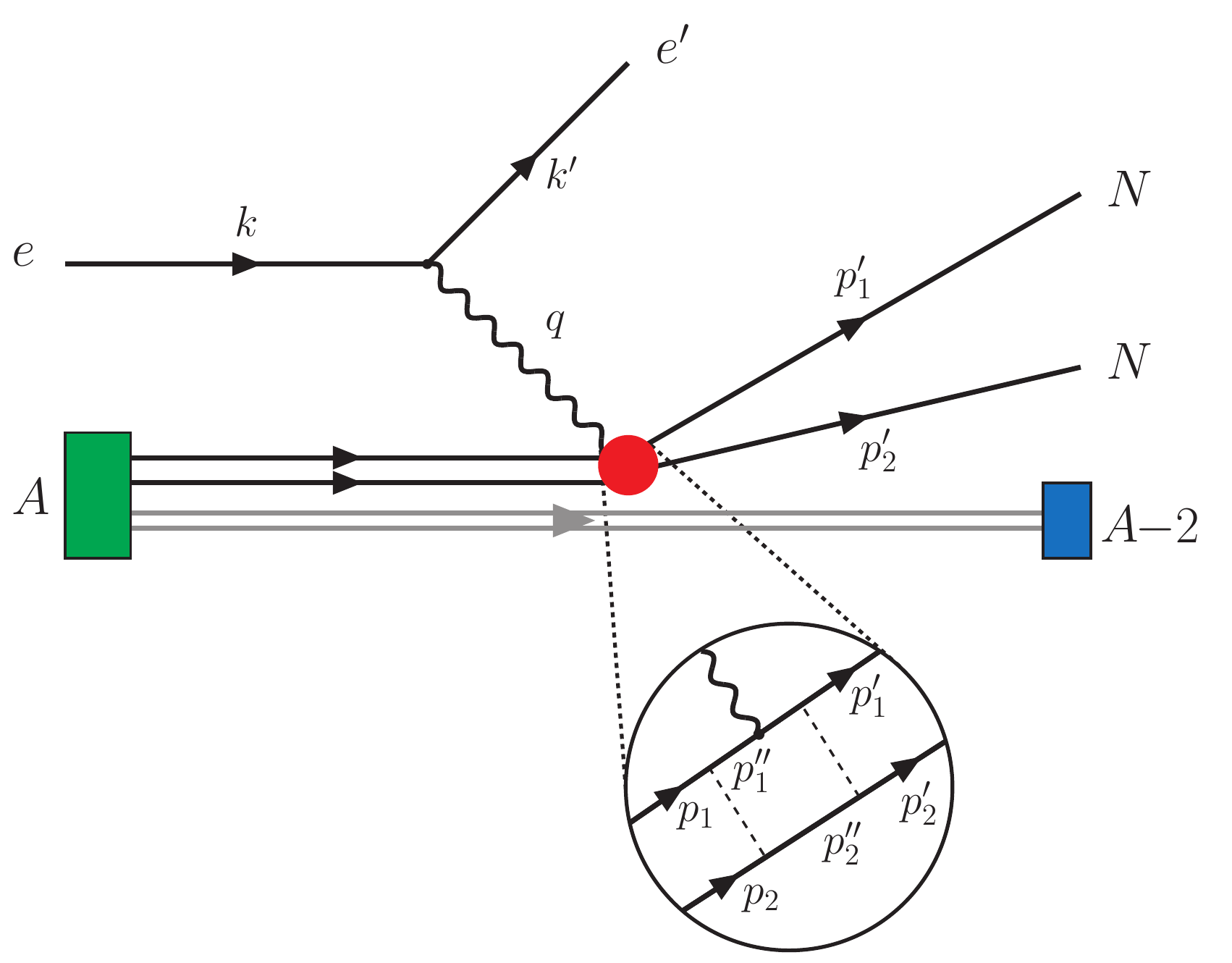}~~~~~%
            \raisebox{.3in}{\includegraphics[width=0.38\textwidth]{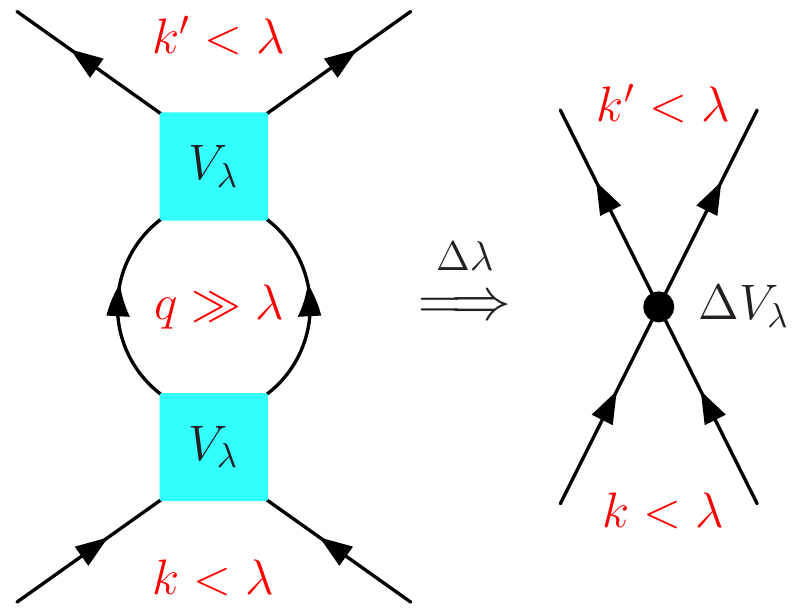}}}
  \vspace*{-.1in} 
  \caption{Schematic two-nucleon knock-out experiment with SRC interpretation (left) and 
  diagrammatic illustration that the contribution
  of decoupled high-momentum modes in intermediate states
  is replaced by (regularized) contact interactions (right). }
  \label{fig:SRCdiagram}      
\end{figure}

Recent electron scattering experiments on nuclei that use large four-momentum
transfers to knock out nucleons
have been interpreted in terms of short-range correlations (SRCs) in the nuclear 
wave function~\cite{Arrington:2011xs,Alvioli:2012qa}.  
As indicated schematically in Fig.~\ref{fig:SRCdiagram} (left), the dominant source
of ejected back-to-back nucleons is identified as the break-up of an SRC formed by
low-momentum nucleons being coupled to high-momentum by the nucleon-nucleon (NN)
interaction.
At the same time, the use of softened (``low-momentum'') Hamiltonians has had great success in pushing the limits
of microscopic calculations of nuclear structure and reactions~\cite{Bogner:2009bt,Roth:2012dv,Furnstahl:2013oba}.
This success is in large part due to the \emph{absence} of SRCs in the corresponding
nuclear wave functions.
We seek to reconcile these results by applying a renormalization group (RG) viewpoint,
which manifests the scale (and scheme) dependence of nuclear Hamiltonians and operators
by continuous changes in the \emph{resolution}.  
RG transformations shift the physics between structure and reaction mechanism 
so that the same data can have apparently different explanations.
We use the RG perspective to discuss implications in light of the current and future
possibilities of applying new RG technology.

The RG is a powerful and versatile tool for this purpose. 
The common features of the RG for critical 
phenomena and high-energy scattering are discussed by Steven Weinberg in an essay in
Ref.~\cite{Guth:1984rq}. He summarizes:
\begin{quote}
``The method in its most general form can I think be understood
as a way to arrange in various theories that the degrees of freedom
that you're talking about are the relevant degrees of freedom for the
problem at hand.''
\end{quote}
This is the essence of what we do by evolving to low-momentum interactions:
we arrange for the degrees of freedom to be the relevant ones for nuclear structure
(and reactions). This does not mean that other degrees of
freedom cannot be used (including SRCs from high-momentum interactions), but we need
to be mindful of Weinberg's adage~\cite{Guth:1984rq}: 
\begin{quote}
``You can use any degrees of freedom you want, but if you use the wrong ones, you'll be sorry.''
\end{quote}
The benefits of applying RG to high-energy (particle) physics include improving perturbation
theory, e.g., in QCD. A mismatch of energy scales can generate large logarithms
that ruins perturbative convergence even when couplings by
themselves are small.
The RG shifts strength between loop integrals and coupling constants
to reduce these logs.  For critical phenomena in condensed matter
systems, the RG reveals the
nature of observed
universal behavior by filtering out short-distance degrees
of freedom.  

Both these aspects are seen in applications
of RG to nuclear structure and reactions.  
As the resolution is lowered, nuclear calculations become more perturbative, implying
that scales are more appropriately matched.
In addition, the potentials flow toward universal form (e.g., see Fig.~\ref{fig:collapse})
as model dependent short-distance details are suppressed.
The end result might be said
to make nuclear physics look more like quantum chemistry calculationally, opening 
the door to a wider variety of techniques (such as many-body perturbation
theory) and simplifying calculations (e.g., by improving convergence
of basis expansions).
However, maintaining RG-induced
three-nucleon (NNN) forces (and possibly four-nucleon forces) 
has been found to be essential for accurate and scale-independent results.
Recently developed RG technology to handle three-body evolution~\cite{Roth:2012dv,Hebeler:2012pr,Wendt:2013bla} 
will be critical to realize the power of the RG.

\begin{figure}[bh]
  \centerline{\includegraphics[width=0.9\textwidth]{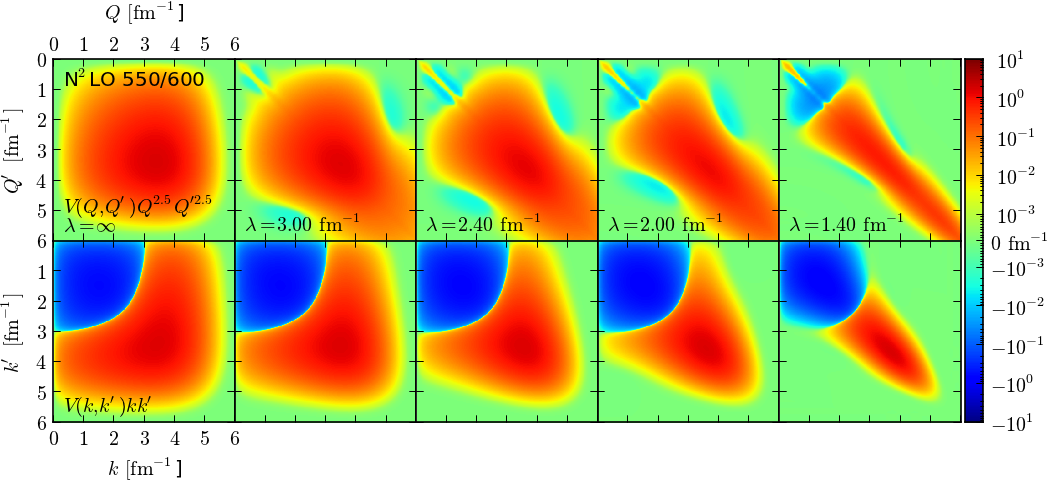}}
  \caption{Evolution by the SRG showing flow toward the momentum diagonal for both NN (bottom)
  and NNN (top) interactions~\cite{Wendt:2013bla}.  The initial potential is an NN + NNN chiral EFT
  interaction at next-to-next-to-leading order (see Ref.~\cite{Epelbaum:2008ga} for background).
  $\lambda$ is a flow parameter with $\hbar^2\lambda^2/M$ roughly equal to the energy decoupling scale.}
  \label{fig:SRGevolution}      
\end{figure}

\section{Similarity renormalization group flow} \label{sec:srg}

Renormalization group methods and applications to nuclear systems 
are well documented in the literature
(see Refs.~\cite{Bogner:2009bt,Furnstahl:2013oba} and references therein) and in
contributions to these proceedings.
A popular approach, which we focus on here, 
is the similarity renormalization group (SRG).
In most implementations of the SRG, an initial Hamiltonian (typically with both NN
and NNN interactions) is driven by a series of continuous unitary transformations toward
more diagonal form in momentum representation.  This flow toward the diagonal
is illustrated for both NN and
NNN matrix elements in Fig.~\ref{fig:SRGevolution}~\cite{Wendt:2013bla}.  
More diagonal means greater decoupling of low- and high-momentum modes, making interactions more
perturbative. 
The changes in many-body interactions highlight the need to be able to control
this part of the evolution.

\begin{figure}[hb!]
  \centerline{\includegraphics[width=0.45\textwidth]{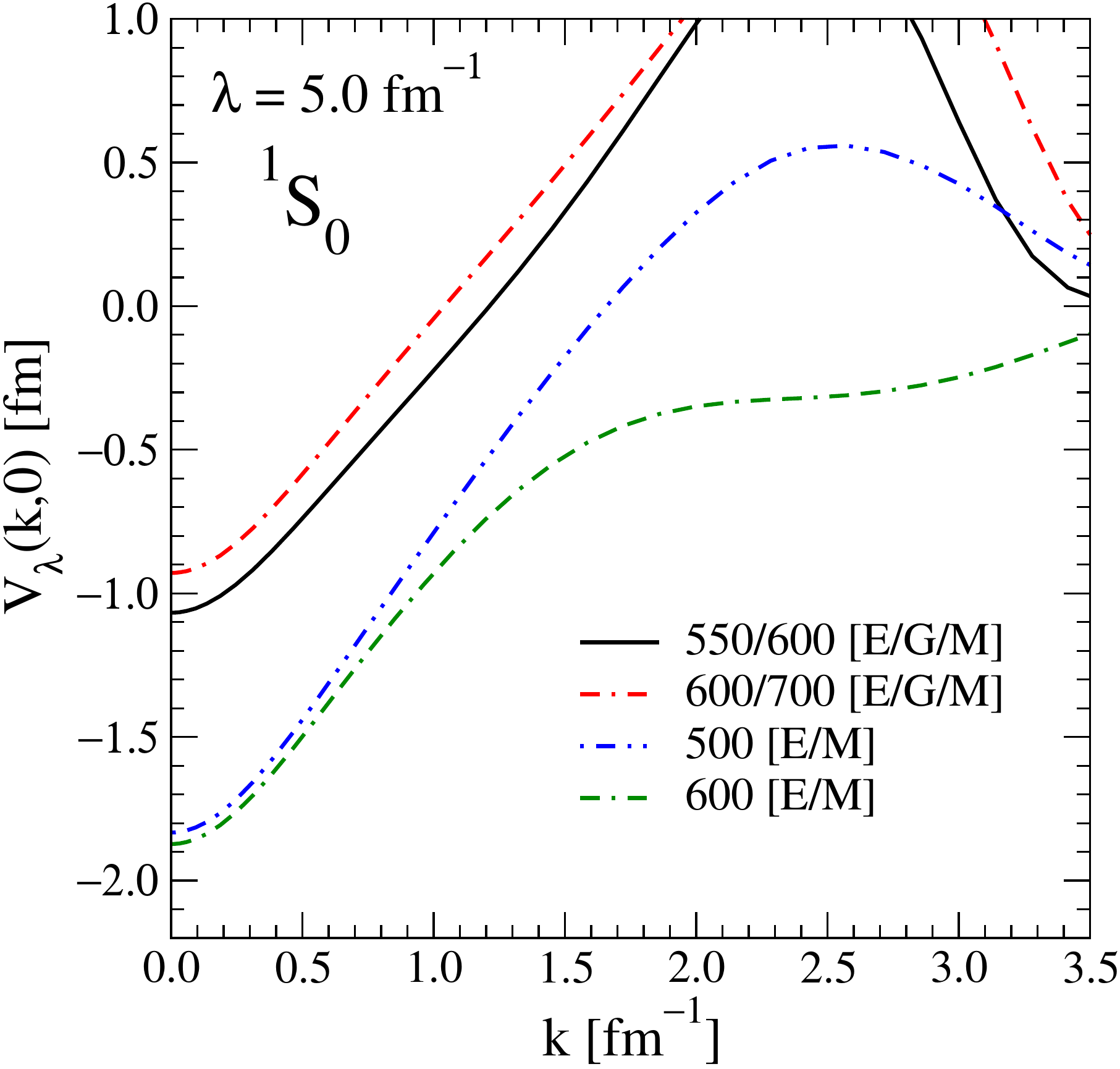}~~~~~%
            \includegraphics[width=0.45\textwidth]{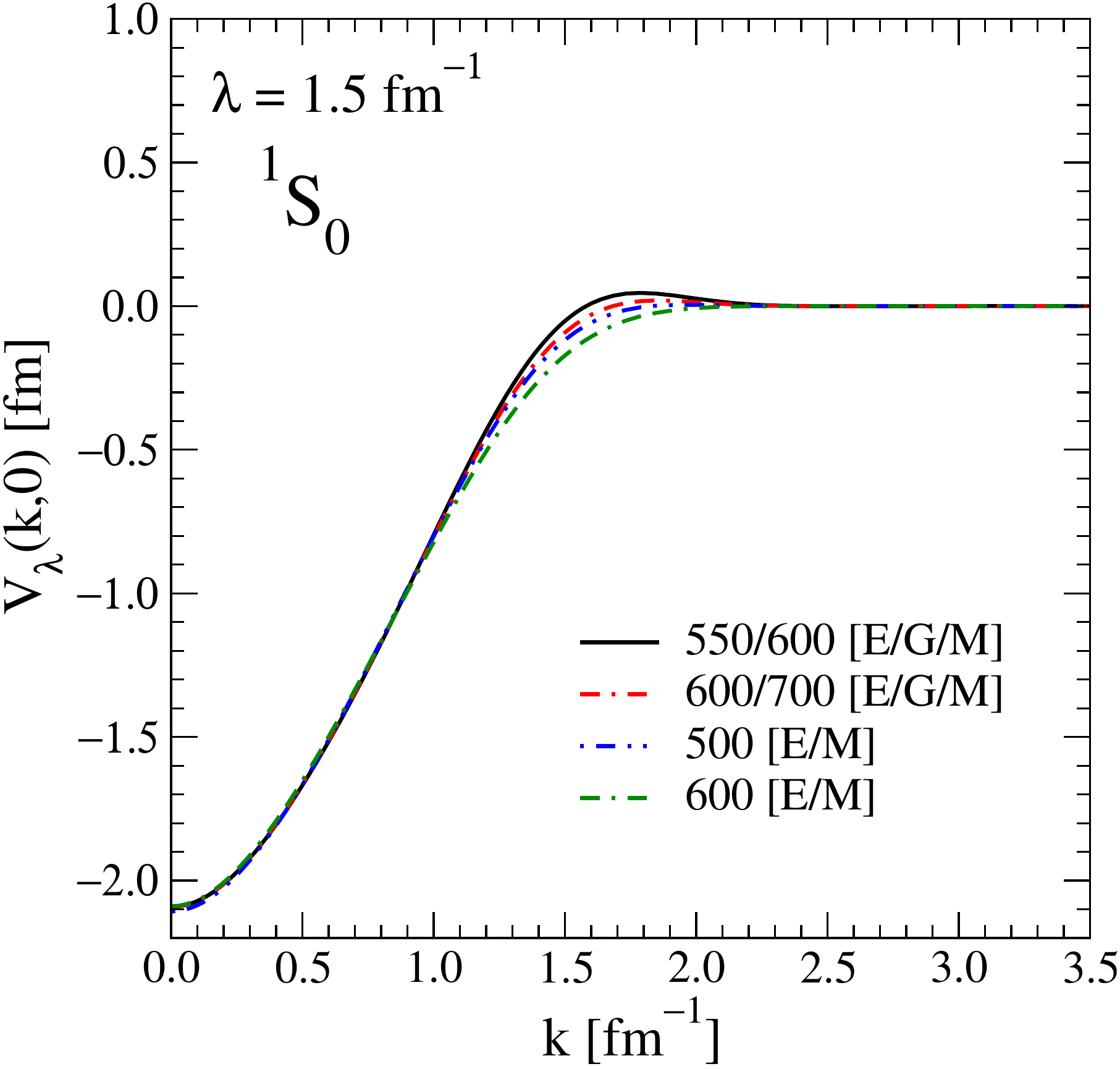}}
  \caption{Off-diagonal matrix elements of different chiral EFT potentials evolved by the SRG
  slightly to $\lambda = 5\fmi$ (left) and much further to $\lambda = 1.5\fmi$ 
  (right)~\cite{Bogner:2009bt}. }
  \label{fig:collapse}      
\end{figure}

Where does the physics of the decoupled high-momentum modes go?
It flows to modifications of the low-momentum parts of both the two- and three-nucleon
interactions, which effective field theory (EFT) tells us can be absorbed
into regulated contact interactions (as indicated schematically
on the right in Fig.~\ref{fig:SRCdiagram}).  That the leading change in the NN
potential induced by the SRG does have this form
can be shown using the operator product expansion 
(see Section~\ref{sec:operators} and Refs.~\cite{Anderson:2010aq,Bogner:2012zm})
but it is implicit in the NN $^1$S$_0$ partial wave in Fig.~\ref{fig:collapse},
where the off-diagonal matrix elements
of a set of chiral EFT NN potentials with different regularization schemes (left) 
are evolved to low resolution (right).
We directly see the suppression of off-diagonal strength for $k > \lambda$
and a flow to universal
values when the high-momentum model dependence is suppressed
(evidence for universal flow has also been observed in three-body
evolution~\cite{Hebeler:2012pr,Wendt:2013bla}).  The dominant change in the potential
at low momentum is a constant shift, as would be expected from changing the strength of
a regulated (smeared) delta function
in coordinate space. 

\begin{figure}[t]
  \centerline{\includegraphics[width=0.99\textwidth]{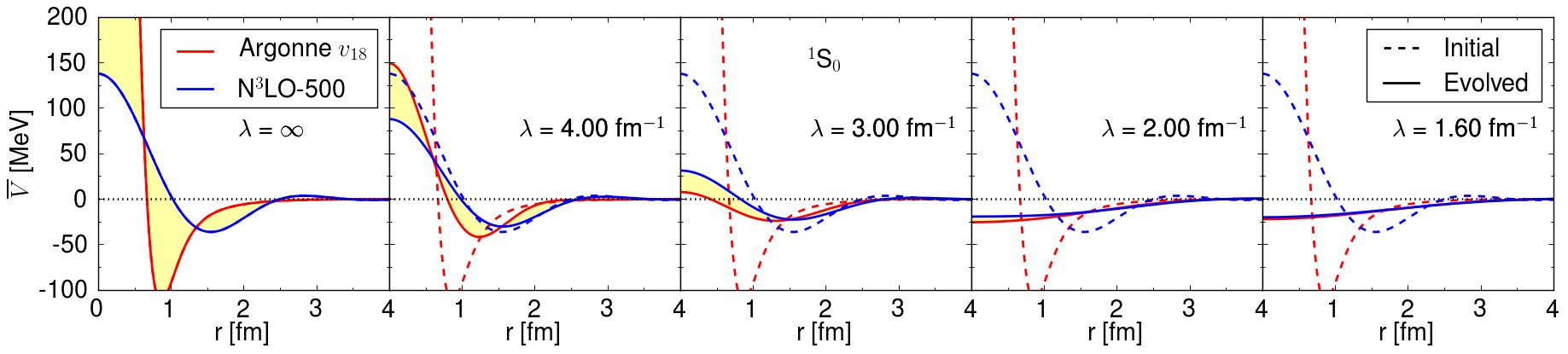}}
  \caption{Evolution by the SRG of two NN interactions in coordinate space as
  visualized by local projections (see Ref.~\cite{Wendt:2012fs} for definitions
  and details). }
  \label{fig:localprojection}      
\end{figure}

\begin{figure}[h]
  \centerline{\includegraphics[width=0.45\textwidth]{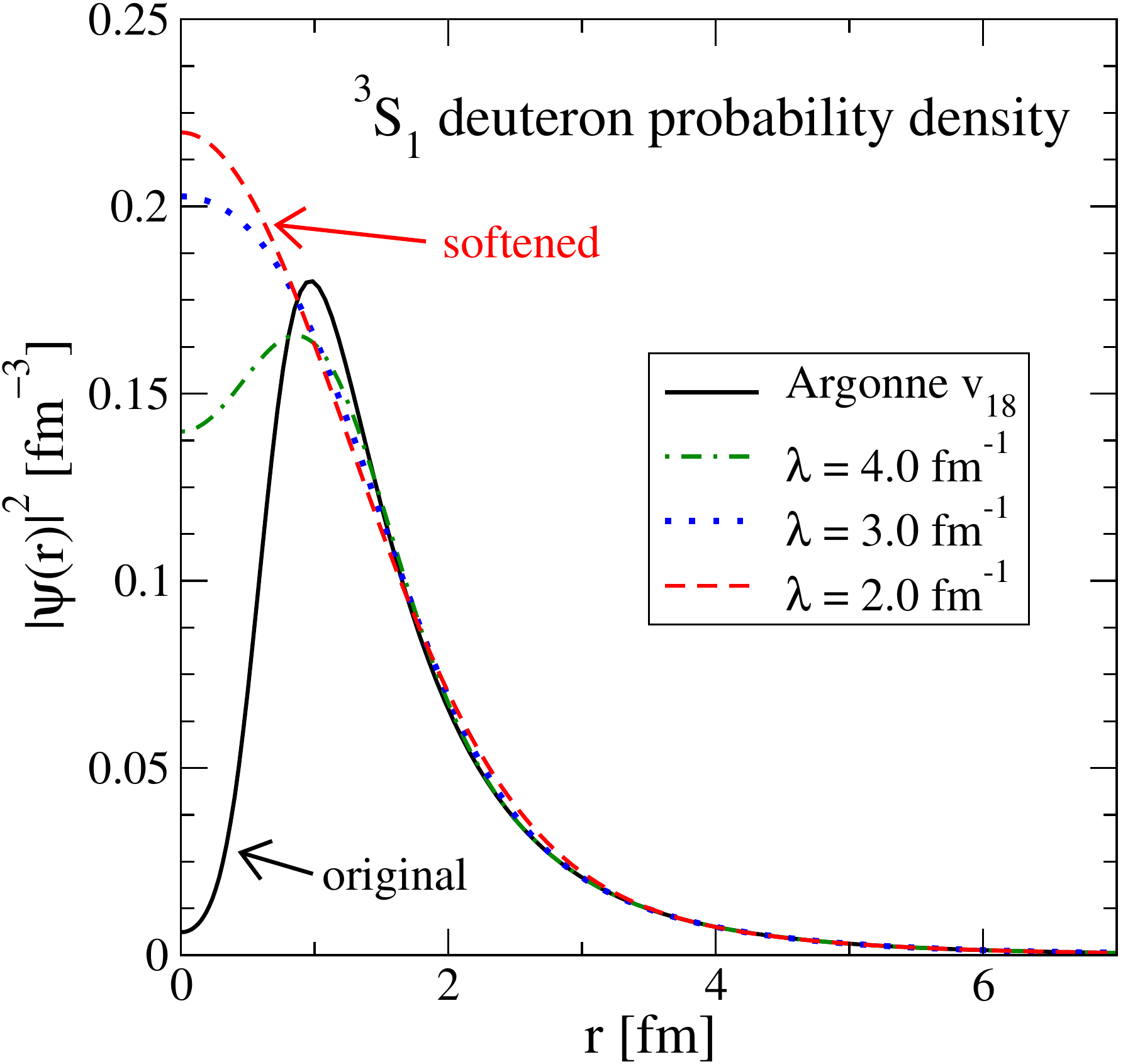}~~~~~%
            \includegraphics[width=0.43\textwidth]{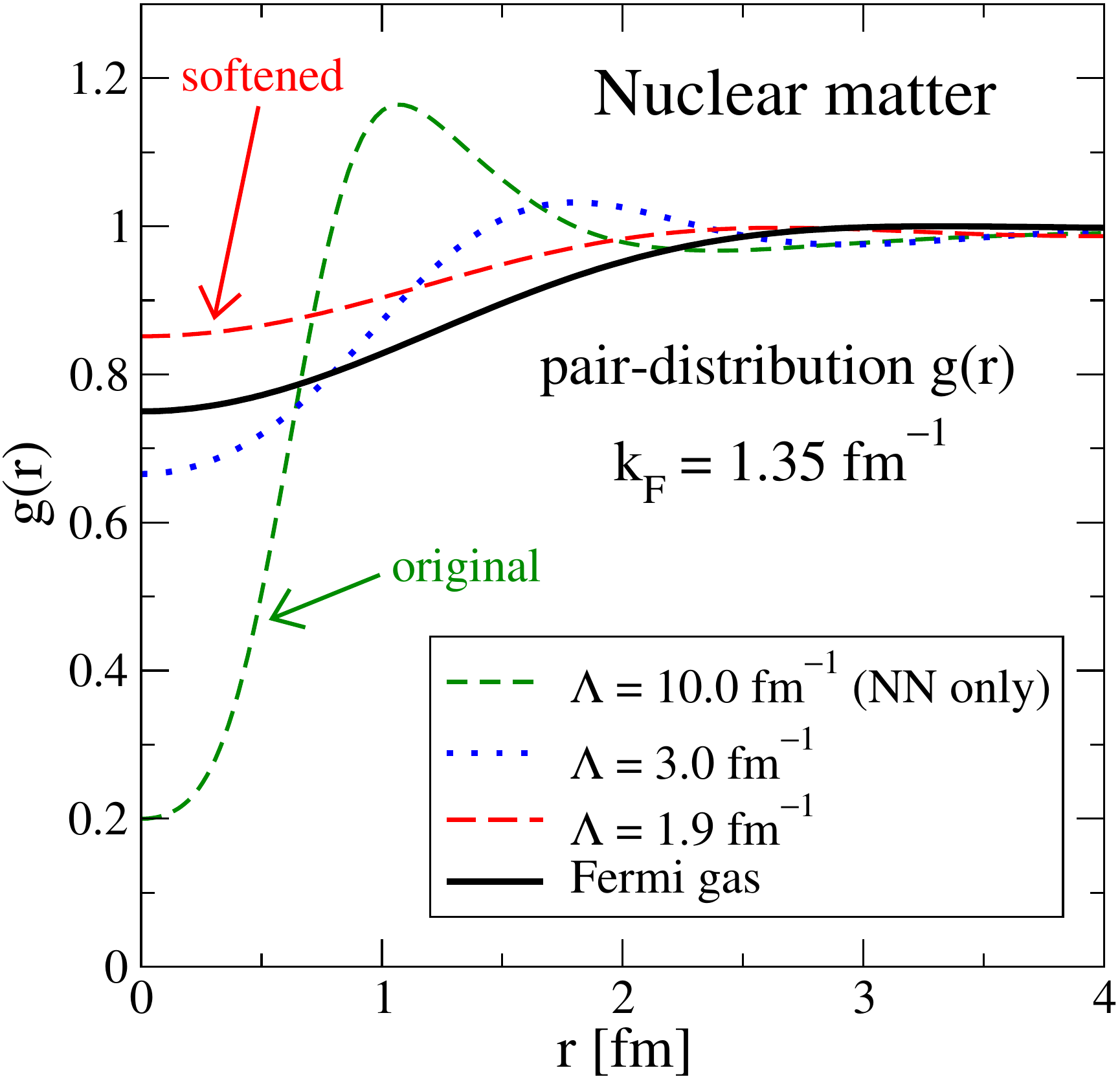}}
  \caption{Short-range correlations induced in the $L=0$ part of the deuteron 
  wave function by the Argonne $v_{18}$ potential~\cite{Wiringa:1994wb}, seen 
  as a suppression at short distances, is removed with SRG evolution to
  $\lambda=2\fmi$ (left). Short-range correlations in nuclear matter, which are manifested
  by the ``wound'' in the pair distribution function compared to the Fermi
  gas, are largely removed by $\vlowk$ RG evolution to $\Lambda= 1.9\fmi$~\cite{Bogner:2009bt}.
  }
  \label{fig:srcs}      
\end{figure}

A visualization of how two-nucleon interactions evolve in coordinate representation
is given for two potentials in the $^1$S$_0$ partial wave in Fig.~\ref{fig:localprojection},
where local projections are applied to the intrinsically non-local SRG evolved
potentials~\cite{Wendt:2012fs}. The melting of the hard repulsive core is manifest
as well as the flow to universal form.  
The soft NN (and NNN) potentials after evolution are much more amenable to
many-body methods that use basis expansions, such as the no-core shell model,
coupled cluster, and the in-medium SRG (see Ref.~\cite{Furnstahl:2013oba} for recent
results).
Indeed, nuclear structure and low-energy reactions are more natural with 
low-momentum interactions, because the Fermi momentum sets the scale rather than a 
repulsive core.
The successes of this approach with the SRG and other RG methods (e.g., which enable
many-body perturbation theory for the shell model) are reviewed
elsewhere~\cite{Bogner:2009bt,Furnstahl:2013oba}.
But because the repulsive core is the
dominant source of SRCs, the nuclear wave functions have variable SRCs as the
resolution is changed (i.e., as $\lambda$ is lowered).
This is illustrated in Fig.~\ref{fig:srcs} for the deuteron (left) and nuclear
matter (right).  What then are the implications for RG evolution for the
high-resolution (that is, high four-momentum transfer) electron scattering experiments?
How does the resolution of the nuclear states even enter the analysis?
To address these questions, we must ask about the evolution of operators
other than the Hamiltonian.

\section{Operator evolution by the SRG} \label{sec:operators}

To gain insight into how RG changes in scale should enter
the analysis of nuclear knock-out experiments, we can use the extraction of parton distribution
functions from deep
inelastic scattering (DIS) as a paradigm.
The key property that make parton distributions well defined
is the controlled factorization of the cross section into structure and reaction
parts at hard scales (meaning sufficiently large $Q^2$)~\cite{Brock:1993sz}.
By this means, a structure function such as $F_2(x,Q^2)$ is decomposed into short-distance
physics from the electron-quark scattering that is captured in Wilson coefficients
in $\widehat F^a_2(x,\frac{Q}{\mu_f})$
and the remainder, which is the soft, long-distance physics defining the parton
distribution $f_a(x,\mu_f)$ (where $a$ labels quarks):
\beq
 F_2(x,Q^2) \sim \sum_a f_a(x,\mu_f) \otimes \widehat F^a_2(x,\frac{Q}{\mu_f})
 \;.
\eeq
The \emph{choice} of the factorization scale $\mu_f$ defines the border between the
long- and short-distance contributions.  It is not unique!  But because the
observable $F_2$ must be independent of $\mu_f$, knowing how the short-distance
part changes with $\mu_f$ determines the RG running of the parton distribution.
A typical choice is $\mu_f = Q$ (to minimize logarithmic contributions to the
Wilson coefficient for
the optimal extraction of PDFs from experiment), so
this running translates into a $Q^2$ dependence in the parton distribution~\cite{Brock:1993sz}.

\begin{figure}[h!]
  \centerline{\includegraphics[width=0.45\textwidth]{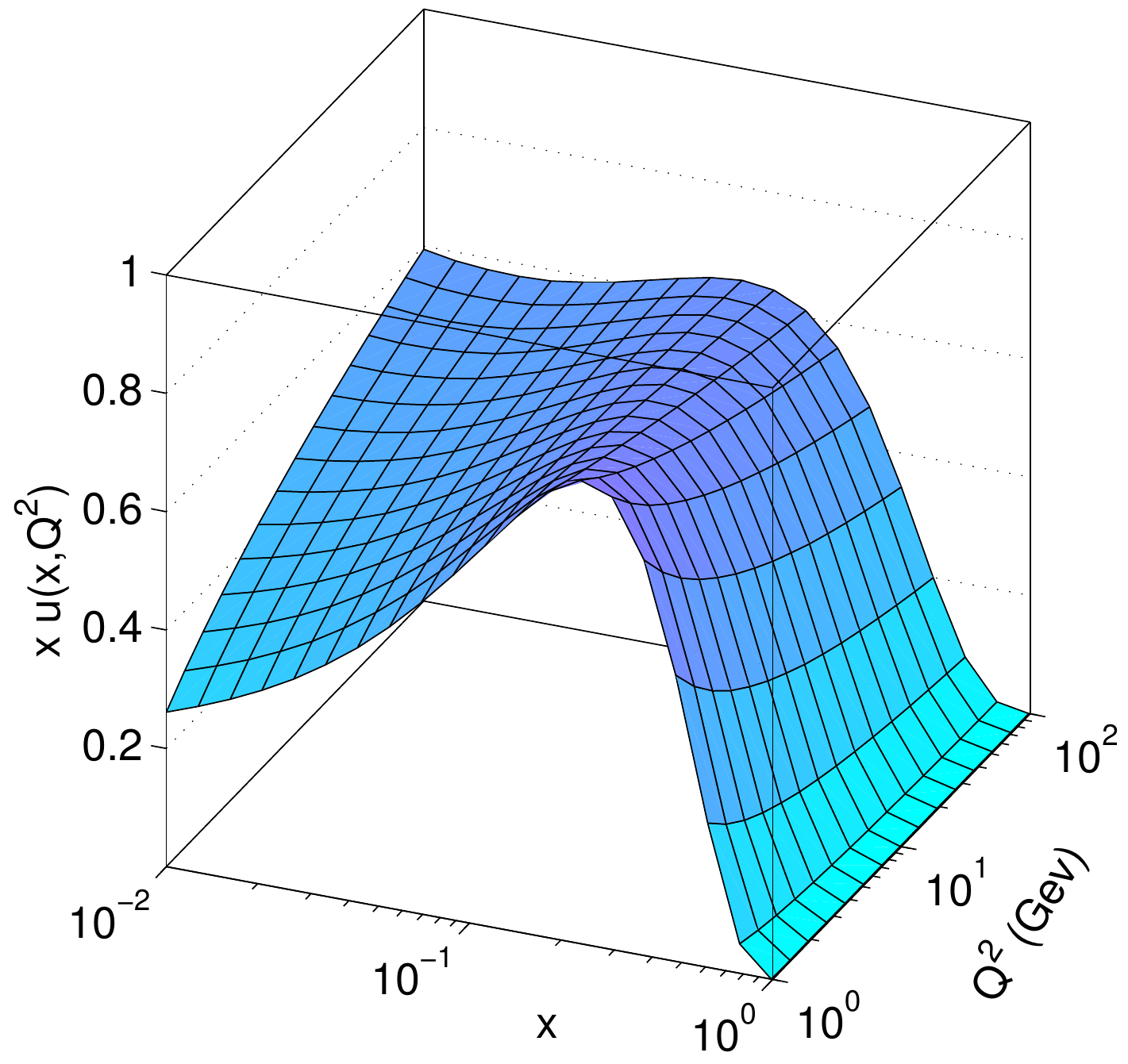}~~~~~%
            \includegraphics[width=0.45\textwidth]{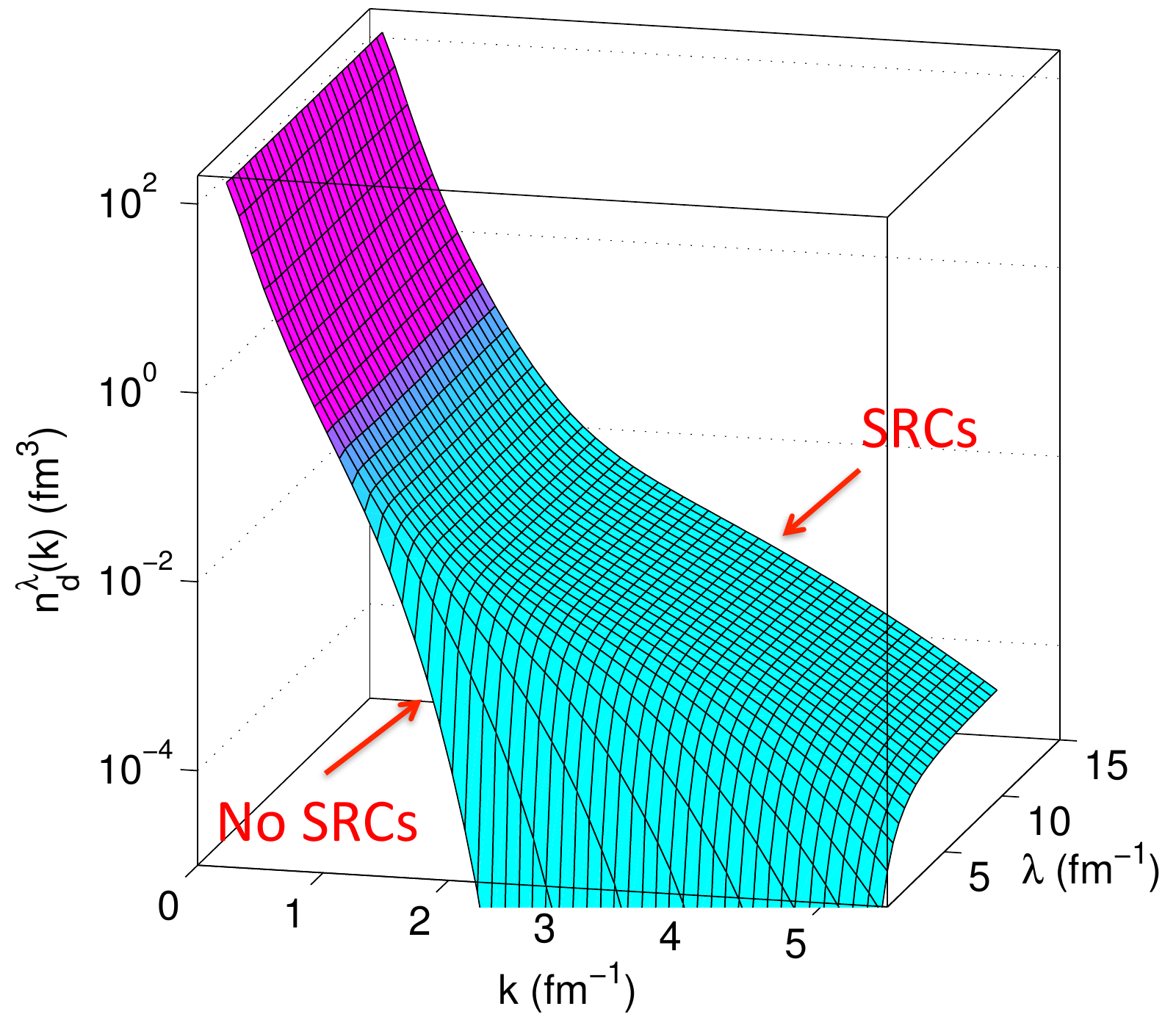}}
  \caption{Parton distribution $x u(x,Q^2)$ for the $u$-quarks in the proton as a function 
   of $x$ and $Q^2$ (left, calculated from~\cite{Lai:1999wy}) and deuteron
   momentum distribution $n^\lambda_d(k)$ at different SRG resolutions $\lambda$ (right).}
  \label{fig:mds}      
\end{figure}

An example of this RG running is shown for the $u$-quark PDF in a proton as a function of
$x$ and $Q^2$ in Fig.~\ref{fig:mds}.
In the left panel, the combination $x u(x,Q^2)$ measures the share of momentum carried by $u$-quarks in a proton within a particular $x$-interval~\cite{Povh:706817,Lai:1999wy}.  This momentum distribution changes as a function
of the resolution scale $Q^2$ according to RG evolution equations.
Thus $u(x,Q^2)$ is scale dependent (as well as scheme dependent, see Ref.~\cite{Brock:1993sz}).
In the right panel, we see that the deuteron momentum distribution $n_d^{\lambda}(k)$
is also scale and scheme dependent.
Plotted is $n^{\lambda}(k)$ for an initial AV18 potential~\cite{Wiringa:1994wb}
(the choice of potential is a \emph{scheme} dependence), which
is  SRG-evolved from $\lambda = \infty$
(corresponding to the initial potential and high resolution) down to $\lambda = 1.5\fmi$
(lowest resolution).
It is evident that the high-momentum tail, which is identified with SRC physics,
is highly scale dependent and is essentially eliminated at lower resolution.

The extraction of momentum distributions or quantities such as spectroscopic factors
from nuclear experiments is also predicated on factorization assumptions just as in DIS.  
That is, the observable
cross sections are separated into the structure and reaction parts according
to some assumptions, which is once again
not a unique decomposition but depends on the factorization scale.  If the impulse
approximation is accurate for some scale, then the separation is clean.  But this
is rarely true in nuclear physics (at least not to the precision we hope to reach).  
Therefore we should ask for the nucleon knock-out experiments the same questions 
that are carefully addressed
in DIS:  Is the factorization robust?  Is it process dependent?  What is necessary
for consistency between structure and reaction models?  What are the trade-offs
between using different scales (and schemes)?

Let's see how the scale dependence like in DIS works out in the language of SRG
unitary transformations.  The measured cross section is a convolution: reaction 
$\!\otimes\!$ structure, but the separate parts are not unique, only the combination.
A (short-range) unitary transformation $\Uop$ leaves matrix elements 
of an operator $\Ohat$ invariant:
\beq
  O_{mn} \equiv \langle \Psi_m | \Ohat | \Psi_n \rangle
       =  \bigl( \langle \Psi_m | \Uop^\dagger \bigr) \,
           \Uop \Ohat \Uop^\dagger \, 
     \bigl( \Uop | \Psi_n \rangle \bigr)
       =  \langle \widetilde \Psi_m | 
           \widetilde O | \widetilde \Psi_n
         \rangle \equiv \widetilde O_{\widetilde m\widetilde n}
         \;.
\eeq
RG unitary transformations change the decoupling scale, which means that the
effective factorization scale (which determines what goes into the operator
and what into the wave function) is changed. 
Note that matrix elements of the operator $\Ohat$ itself between the transformed
states are in general modified:
\beq
  O_{\widetilde m\widetilde n} \equiv 
  \langle \widetilde\Psi_m | \Ohat  | \widetilde \Psi_n \rangle
  \neq O_{mn}
  \quad
  \Longrightarrow
  \quad
  \mbox{e.g.,}\ 
  \langle \Psi^{A-1}_n | a_\alpha | \Psi_0^A \rangle \ \mbox{changes,} 
\eeq
where the latter is a spectroscopic factor.
In a low-energy effective theory, transformations that modify short-range
unresolved physics yield equally valid states,  so matrix
elements such as spectroscopic factors (SFs)
or momentum distributions (see Fig.~\ref{fig:mds}) are scale/scheme dependent observables.

All ingredients for the analysis of an experimental cross section mix under a 
unitary transformation that changes the resolution.  A one-body current becomes
a many-body current:
\begin{center}    
   \raisebox{.32in}{$ {\Uop\wh\rho(\qvec)\Uopdag}\ =\ $}
     \includegraphics[width=.8in]{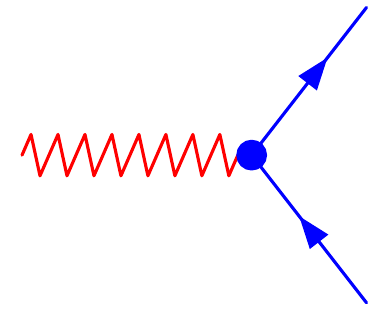}
   \raisebox{.32in}{$\ + \ \alpha $}
     \includegraphics[width=.7in]{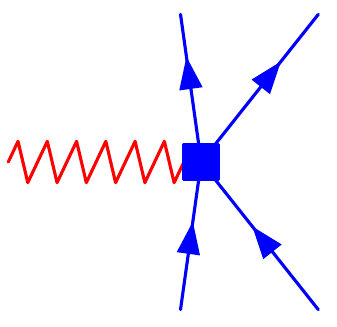}
    \raisebox{.32in}{$\ + \  \cdots \;,$}
\end{center}
final-state interactions are modified, 
and new wave function correlations appear (or disappear in the case of
short-range calculations at lower resolution):
\begin{center}
 \raisebox{.3in}{${ \Uop |\Psi_0^A \rangle} = \Uop$}   
  \includegraphics[width=0.7in]{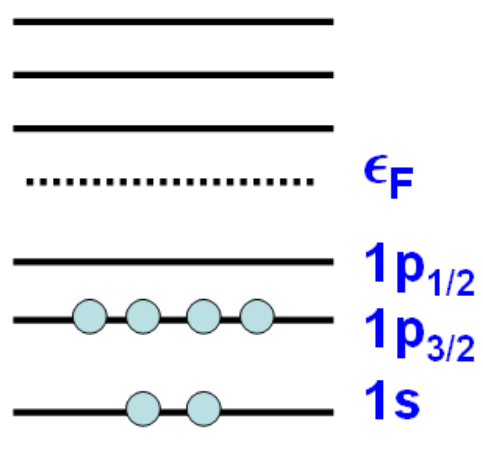}
   \raisebox{.3in}{$\ + \cdots \ \Longrightarrow \ Z$}
  \includegraphics[width=0.7in]{cioffi_ipm_levels}
   \raisebox{.3in}{$\  + \ \alpha $}
  \includegraphics[width=0.72in]{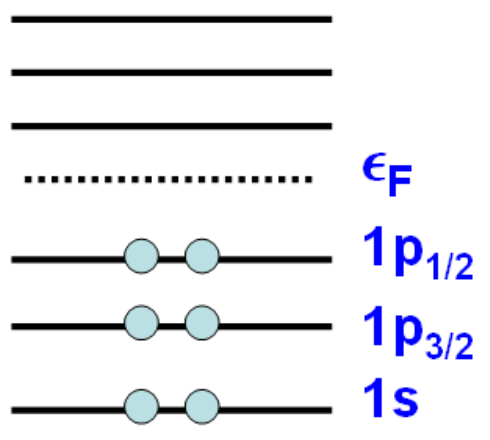}
    \raisebox{.3in}{$\ + \  \cdots $}
\end{center}
Again, this means that quantities such as SFs are scale dependent.
The bottom line is that the cross section is unchanged only if all pieces
are included with the same $U$:  the Hamiltonian, the current operator,
and the final state interactions.

\begin{figure}[t]
  \centerline{\includegraphics[width=0.46\textwidth]{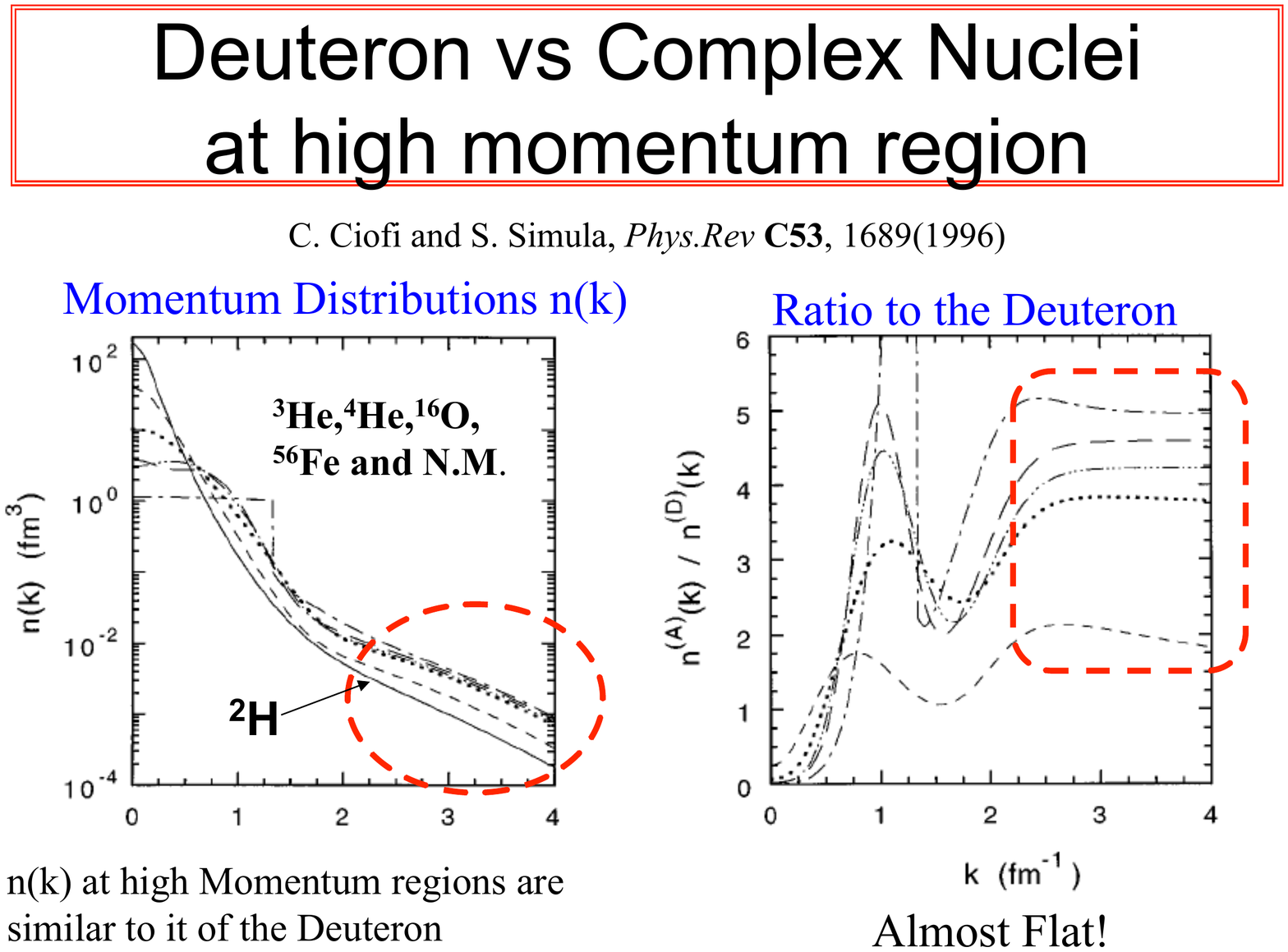}~~~~%
            \includegraphics[width=0.44\textwidth]{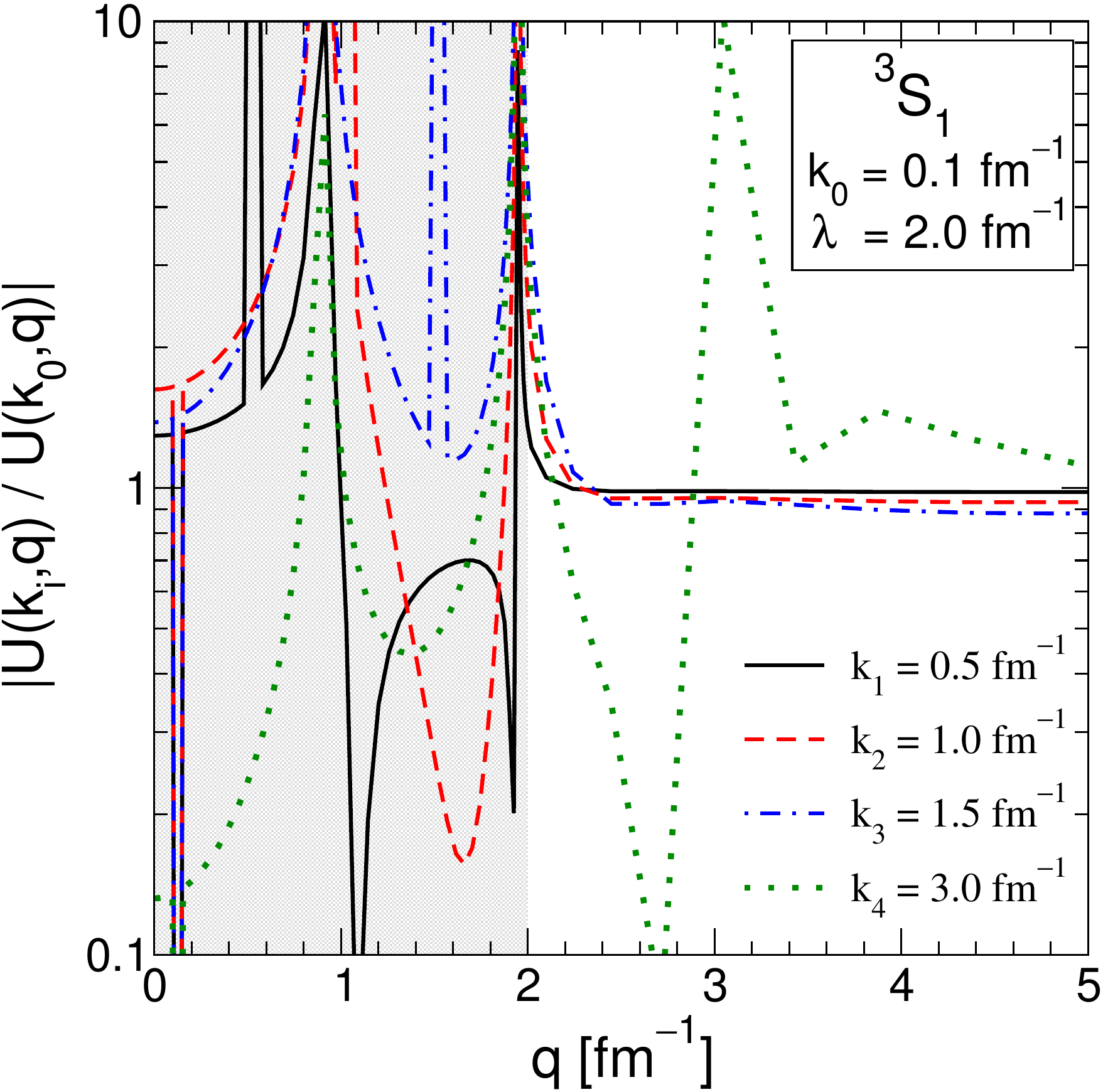}}
  \caption{Ratio of momentum distributions in nuclei to the deuteron
  (denoted $n_A$ and $n_d$ in the text) with a high-resolution
  potential~\cite{CiofidegliAtti:1995qe} (left) and $U_\lambda$-factorization test
  for the $^3$S$_1$ channel~\cite{Anderson:2010aq} (right).}
  \label{fig:ratios}      
\end{figure}

Now consider again the high resolution experiment from Fig.~\ref{fig:SRCdiagram}
and what happens when RG unitary transformations act to change the resolution.
In particular,
how does
the SRC explanation of nuclear scaling, which accounts for plateaus in inclusive
cross section ratios, evolve with the resolution scale?
This explanation is based
on the dominant role played by the one-body current, the two-body interaction, and SRCs.  
The underlying physics is most simply isolated by considering 
the high-resolution momentum distributions in nuclei.
In Fig.~\ref{fig:ratios} on the left, we see that ratios of the 
these momentum distributions
in various nuclei to that in the deuteron are almost flat in the high momentum
region associated with SRCs (i.e., above $k = 2\fmi$).  
The contribution highlighted in the circle in Fig.~\ref{fig:SRCdiagram} yields
a $k$ dependence largely independent of the nuclear environment, so 
$n_A(k)$ simply scales with $A$.
If we now evolve to lower momentum through unitary transformations, this can no
longer explain the cross section, because the softening of the interaction
and therefore the wave function means the momentum distribution has no
support at these high momenta (e.g., as in Fig.~\ref{fig:mds} for the deuteron).

But the cross section \emph{must} be unchanged, because it is a unitary transformation.
With RG evolution, the probability of high momentum in a nucleus decreases, but if
we transform the wave functions \emph{and} operators:
\beq
n(k) \equiv \langle A | \adag_{\kvec} \ap_{\kvec} | A \rangle 
       = {\bigl( \langle A | \Uwh^\dagger \bigr)} \,
          { \Uwh \adag_{\kvec} \ap_{\kvec} \Uwh^\dagger} \, 
    { \bigl( \Uwh | \Psi_n \rangle \bigr)}
       = { \langle \widetilde A | }
         { \Uwh \adag_{\kvec} \ap_{\kvec} \Uwh^\dagger} { | \wt A
         \rangle} \;,
\eeq
then the original momentum distribution is unchanged! 
We know that the transformed state ${ | \wt A \rangle}$ is easier
to calculate, but is the new operator too difficult to calculate
or even pathological (e.g., does it explode to compensate for the super-exponential
suppression of the low-resolution momentum distribution)?

\begin{figure}[t!]
 \centerline{\includegraphics[width=0.85\textwidth]{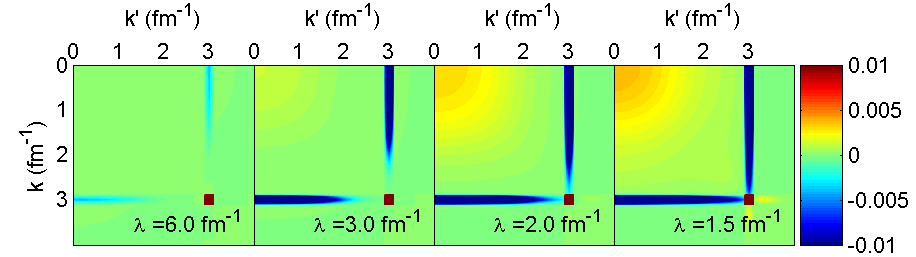}}
 \medskip
 \centerline{\includegraphics[width=0.85\textwidth]{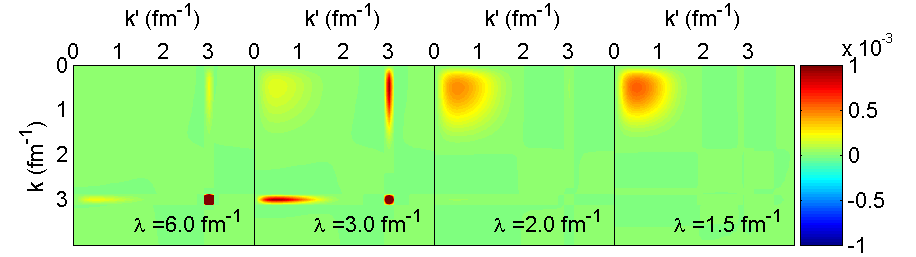}}
  \caption{Integrand of the deuteron momentum distribution at $q\approx 3\fmi$ 
  without (top) and with (bottom) the
  deuteron wave functions included~\cite{Anderson:2010aq}. }
  \label{fig:md_operators}
\end{figure}

Let us consider the SRG operator flow for the momentum distribution
graphically.  
The evolution with $\flow$ of any operator $\Ohat$ is given by: 
\beq
     \Ohat_\flow = \Uop_\flow \Ohat \Uop^\dagger_\flow \;,
\eeq
which can be carried out by a flow equation similar to that used to evolve
the Hamiltonian.  In practice it is more efficient to construct
the unitary transformation 
from $\Uop_\lambda = \sum_i |\psi_i(\lambda)\ra \la \psi_i(0)|$  or
by solving the $dU_\lambda/d\lambda$ flow equation.
In any case, matrix elements of evolved operators are
unchanged by construction (for the deuteron) \emph{but the distribution of strength flows}.
The integrand of the momentum distribution $\langle\psi_d| a^\dagger_q a_q |\psi_d\rangle $
in the deuteron at $q \approx 3.0\fmi$
is shown in Fig.~\ref{fig:md_operators}.
In the top figure, the initial integrand of $\Uop_\lambda a^\dagger_q a_q \Uop_\lambda^\dagger$  
at $\lambda = \infty$ has a delta function at $k = k' = q$.  
In the SRG flow, one-body
operators such as $a^\dagger_q a_q$ do not evolve, and their contribution is in fact unchanged
with $\lambda$.  However, there is a clear flow to lower momentum, which must be
entirely due to a two-body operator.
In the bottom figure, the deuteron wave functions are folded in (such that the
integrated area is the invariant value of the original momentum distribution
at $q=3\fmi$).  We see 
there is negligible amplitude at small $\lambda$ from the original one-body operator
(nothing explodes!),
but instead a smooth contribution at low momenta from the induced two-body
operator, which is reminiscent of a regularized delta function. 

We might wish to conclude that this operator flow implies a type of ``conservation
of difficulty'' with the simplification of the wave function countered by the
complication of the operator.  But in this situation the separation of momentum scales
leads to an important generic factorization of the unitary transformation
operator $U_\lambda$.  In particular, 
$U_\lambda$-factorization says that the two-body unitary transformation becomes
a simple product (in each partial wave):  
 $ U_{\lambda}(k,q)\rightarrow {K_\lambda(k)}{Q_\lambda(q)}$
 whenever $k<\lambda$ and $q\gg\lambda$.  This result follows from 
 applying effective interaction methods or the
operator product expansion (OPE) for nonrelativistic wavefunctions; 
we refer the reader to Refs.~\cite{Anderson:2010aq,Bogner:2012zm} for the 
technical details.  Here we rely on a visual demonstration.
In particular, we test $U_\lambda$-factorization by considering the ratio
of $U_\lambda(k,q)$ at fixed $q$ but variable $k$.
In the factorization region: 
\beq
\frac{U_{\lambda}({ k_{i}},{q})}{U_{\lambda}({ k_{0}},{q})}
  \mathrel{\mathop{\longrightarrow}^{k<\lambda\strut}_{q\gg\lambda}}
  \frac{{K_\lambda(k_{i})}{Q_\lambda(q)}}{{K_\lambda(k_{0})}{Q_\lambda(q)}}
  = \frac{K_\lambda(k_{i})} {K_\lambda(k_{0})}
  \approx 1 \;,
\eeq
so for $q\gg\lambda$ we expect the ratio to go to a constant, which is in fact
unity because $K_\lambda(k)$ becomes independent of $k$ to leading order in the OPE.
In Fig.~\ref{fig:ratios} (right), we plot this ratio in the
$^3$S$_1$ channel and see clear plateaus close to one (at the 10--15\% level)
for those curves with $k_i < \lambda$ in the $q>\lambda$ region, just as expected.
It works similarly in other channels~\cite{Anderson:2010aq}.

We emphasize that because the leading order for $K_\lambda(k)$ is constant
for $k < \lambda$,
the factors $K_\lambda(k)K_\lambda(k')$ to good approximation 
play the role of a contact term.
Then the contribution from large $\lambda$ in the diagram in Fig.~\ref{fig:SRCdiagram} (right) 
with an implied integration over $q$ and $q'$ has the simplification:
 \begin{align}
    \Delta V_{\lambda}(\kl,\kpl)
    &= \int_{q,q'} U_\lambda(\kl,\qh) {V_{\lambda}(\qh,\qph)} U^\dagger_\lambda(\qph,\kpl)
      \ \mbox{for\ } { \kl,\kpl <\lambda},\ {\qh,\qph\gg\lambda} \nonumber \\
    &\stackrel{U_\lambda\rightarrow {K}\cdot {Q}}{\longrightarrow}
      K(k) \left[\int_{q,q'} Q(q) V_{\lambda}(q,q')Q(q')\right ] K(k') \ \mbox{with\  } K(k) \approx 1
      \;,
 \end{align}  
which is a constant times a smeared delta function, as advertised.
Further, we can understand why nuclear scaling is expected 
directly from $U_\lambda$-factorization, if we can argue that the deuteron channel
dominates (as in the SRC argument~\cite{Arrington:2011xs,Alvioli:2012qa}).
When $k < \lambda$ and $q \gg\lambda$, the ratio of original momentum distributions
becomes (in a schematic notation):
\begin{align}
  \frac{n_A(q)}{n_d(q)} 
       &= \frac
         {\langle \wt A | \Uwh \adag_{\qvec} \ap_{\qvec} \Uwh^\dagger | \wt A \rangle}
         {\langle \wt d | \Uwh \adag_{\qvec} \ap_{\qvec} \Uwh^\dagger | \wt d \rangle}
    =
        \frac
       {\langle \wt A | \int\! U_{\lambda}(k',q') \delta_{q'q} U^\dagger_{\lambda}(q,k)
         | \wt A \rangle}
       {\langle \wt d | \int\! U_{\lambda}(k',q') \delta_{q'q} U^\dagger_{\lambda}(q,k) 
         | \wt d \rangle} 
  \nonumber \\
  &=
        \frac
       {\langle \wt A | \int\! K_\lambda( k') [\int\! Q_\lambda(q')  \delta_{q'q} Q_\lambda(q)]
         K_\lambda(k)  | \wt A \rangle 
       }
       {\langle \wt d | \int\! K_\lambda( k') [\int\! Q_\lambda(q')  \delta_{q'q} Q_\lambda(q)]
         K_\lambda(k) | \wt d \rangle
       } 
  =
\frac{\langle \wt A | \int\! K_\lambda( k') K_\lambda(k) | \wt A \rangle}
     { \langle \wt d | \int\! K_\lambda( k') K_\lambda(k) | \wt d \rangle} 
   \nonumber \\
 &\equiv C_A
 \;,
  \label{eq:ratiomds}
\end{align}
where $C_A$ is the scaling ratio.  A proof of principle test in a toy one-dimensional
model verified that this scenario can work~\cite{Anderson:2010aq}.  For the realistic 
nuclear case, we need to examine all contributions quantitatively, including from
three-body operators, but the pattern in Eq.~\eqref{eq:ratiomds} is promising.

We might further speculate that the recent observation that
the $A$ dependences of scaling ratios and the slope of the 
EMC effect $dR_A(x)/dx$ (where $R_A(x)$
is the large $Q^2$ ratio of nuclear cross sections for $0.7 < x < 1.0$) are linearly
correlated~\cite{Arrington:2011xs} could be understood by $U_\lambda$-factorization and subsequent cancellations in cross section ratios.
The EFT treatment of Chen and Detmold~\cite{Chen:2004zx} predict an
analogous factorization
in the EMC ratio.  In particular, they assert: 
\begin{quote}
 ``The $x$ dependence of $R_A(x)$ is governed by short-distance physics, while the overall magnitude (the $A$ dependence) of the EMC effect is governed by long distance matrix elements calculable using traditional nuclear physics.''
\end{quote}
If the same leading operators dominate in the two types of processes
(i.e., two-body contact operators with deuteron
quantum numbers), then we would expect precisely this sort
of linear $A$ dependence.  Quantitative calculations are needed!
 
To do such calculations, we need many-body operator contributions, as shown by 
Neff for $^4$He relative momentum distributions~\cite{Neff:2013private}.
Fortunately, the recently developed technology for evolution of three-body forces
can be adopted for more general operator evolution.  This will enable direct calculations
by \textit{ab initio} methods 
in lighter nuclei and
many-body perturbation theory for operators in heavier nuclei.

\section{Summary and outlook}  \label{sec:summary}

We have presented a brief overview of high-resolution probes of low-resolution
nuclei based on the RG/EFT perspective.
Some summary observations: 
\begin{itemize}
      \I Lower resolution means more natural nuclear structure.
      \I While scale and scheme-dependent observables can be (to good
      approximation) unambiguous for
          \emph{some} systems, they are often (generally?) not so for nuclei.
    And while cross sections are invariant, the physics interpretation
    can change with resolution!
     \I Working with scale and scheme dependence
     requires \emph{consistent} Hamiltonian and operators. 
       Be wary of treating experimental analysis in independent pieces (as 
       is often done).
     \I Unitary transformations can be used to reveal \emph{natural} scheme dependence.
\end{itemize}
The RG/EFT perspective and associated tools can help to address whether we can have
controlled factorization at low energies, to identify the roles of short-range
versus long-range correlations, and to quantitatively assess the scheme-dependence
of spectroscopic factors and related quantities.

An overreaching question is
how should one choose the appropriate scale in different situations (with the RG
to evolve the scale as needed).  One general motivation is to make calculations
easier or more convergent, such as using the QCD running-coupling scale to improve perturbation
theory.  For nuclear structure and low-energy reactions,
low-momentum potentials are chosen to improve convergence
in configuration interaction or coupled cluster calculations or to make a microscopic
connection to the shell model.  Conversely, local potentials (which until recently
were only high resolution) are favored for quantum Monte 
Carlo.  The scale could also be chosen for interpretation or intuition; the SRC phenomenology
is such an example.  
But the most important issue for knock-out experiments is to have
the cleanest and most controlled extraction of quantities analogous to PDFs from experiment; 
this might mean optimizing the validity of the impulse approximation but there are other
possibilities (e.g., optimizing $U_\lambda$-factorization).  
To make progress, the plan is to make test calculations with a range
of scales starting from initial Hamiltonians and operators
matched in an EFT framework, with the RG used to consistently
relate scales and quantitatively probe ambiguities (e.g., in spectroscopic factors).
A priority calculation in the short term
is deuteron electrodisintegration, which is well controlled
because of the absence of three-body forces and operators. 

\vspace*{.1in}

I thank my collaborators
E. Anderson, S. Bogner, B. Dainton,
K. Hebeler, H. Hergert, S. More, R. Perry, A. Schwenk, and K. Wendt.
This work was supported in part by the National Science Foundation 
under Grant No.~PHY--1002478 and
the U.S.\ Department of Energy under Grant No.~DE-SC0008533 (SciDAC-3/NUCLEI project).


\bibliographystyle{h-physrev4}
\bibliography{srg_refs}

\end{document}